\documentclass{article}

\usepackage[preprint,nonatbib]{neurips_2020}

\usepackage[numbers]{natbib}

\usepackage[utf8]{inputenc} % allow utf-8 input
\usepackage[T1]{fontenc}    % use 8-bit T1 fonts
\usepackage{hyperref}       % hyperlinks
\usepackage{url}            % simple URL typesetting
\usepackage{booktabs}       % professional-quality tables
\usepackage{amsfonts}       % blackboard math symbols
\usepackage{nicefrac}       % compact symbols for 1/2, etc.
\usepackage{microtype}      % microtypography

\usepackage{amsmath}
\usepackage{bbm}
\usepackage{graphicx}

\DeclareMathOperator*{\argmin}{arg\,min}
\DeclareMathOperator*{\argmax}{arg\,max}
\DeclareMathOperator{\asinh}{sinh^{-1}}

\usepackage{xcolor, soul}
\definecolor{orange}{rgb}{1,0.5,0}
\definecolor{mydarkcyan}{rgb}{0,0.5,0.5}

\title{Deep Potential: Recovering the gravitational potential from a snapshot of phase space}

\author{%
  Gregory M. Green\\
  Max Planck Institute for Astronomy\\
  D-69117 Heidelberg, Germany\\
  \texttt{green@mpia.de}\\
  \And
  Yuan-Sen Ting\\
  Institute for Advanced Study, Princeton\\
  Princeton, NJ 08540, United States\\
  \texttt{ting@ias.edu} \\
}

\begin{document}

\maketitle

\begin{abstract}
  One of the major goals of the field of Milky Way dynamics is to recover the gravitational potential field. Mapping the potential would allow us to determine the spatial distribution of matter -- both baryonic and dark -- throughout the Galaxy. We present a novel method for determining the gravitational field from a snapshot of the phase-space positions of stars, based only on minimal physical assumptions. We first train a normalizing flow on a sample of observed phase-space positions, obtaining a smooth, differentiable approximation of the phase-space distribution function. Using the collisionless Boltzmann equation, we then find the gravitational potential -- represented by a feed-forward neural network -- that renders this distribution function stationary. This method is far more flexible than previous parametric methods, which fit narrow classes of analytic models to the data. This is a promising approach to uncovering the density structure of the Milky Way, using rich datasets of stellar kinematics that will soon become available.
\end{abstract}

\section{Introduction}

To know the gravitational potential of the Milky Way is to know the three-dimensional distribution of matter. Stars and gas make up most of the baryonic mass of the Galaxy. However, dark matter is only detectable through its gravitational influence. Mapping the gravitational potential in 3D is therefore key to mapping the distribution of matter -- both baryonic and dark -- throughout the Galaxy.

The trajectories of stars orbiting in the Milky Way are guided by gravitational forces. Were it possible to directly measure accelerations of individual stars due to the Galaxy's gravitational field, then each star's acceleration would indicate the local gradient of the gravitational potential \citep{Quercellini2008,Ravi2019}. This would allow us to use Hamiltonian neural network approaches to learn the gravitational potential \citep{Greydanus2019}. However, the scale of these gravitational accelerations -- on the order of $1\,\mathrm{cm\,s^{-1}\,yr^{-1}}$ -- is beyond the ability of current spectroscopic and astrometric instruments to measure \citep{Silverwood2019}. We instead observe a frozen snapshot of stellar positions and velocities. The gravitational potential determines how the phase-space density, the ``distribution function,'' evolves in time. Unless one invokes further assumptions, any gravitational potential is consistent with any snapshot of the distribution function, as the potential only determines the \textit{time evolution} of the distribution function. A critical assumption of most dynamical modeling of the Milky Way is therefore that the Galaxy is in a steady state, meaning that its distribution function does not vary in time \citep{Binney2013,BlandHawthornGerhard2016}.

State-of-the-art dynamics modeling techniques generally work with simplified analytic models of the distribution function and gravitational potential. The results produced by such techniques can only be as good as the models that are assumed. This motivates us to go beyond simple parametric models. Here, we demonstrate a technique that learns highly flexible representations of both the distribution function and potential. Our method makes only minimal assumptions about the underlying physics:
\begin{enumerate}
    \item Stars orbit in a time-independent gravitational potential $\Phi \left( \vec{x} \right)$.
    \item We have observed the phase-space coordinates of a population of stars that are statistically stationary (\textit{i.e.}, whose phase-space distribution does not change in time).
    \item The gravitational potential is related to the matter density by Poisson's equation: ${\nabla^2 \Phi = 4\pi G \rho \left( \vec{x} \right)}$. Matter density is non-negative everywhere: ${\rho \left( \vec{x} \right) \geq 0}$. Thus, ${\nabla^2 \Phi \geq 0}$.
\end{enumerate}
We represent the distribution function using a normalizing flow, and the gravitational potential using a densely connected feed-forward neural network. We train the normalizing flow to represent the distribution of the observed phase-space coordinates of the stars, and then train the gravitational potential to render this distribution stationary, subject to the constraint that matter density must be positive. We thus use highly flexible representations for the distribution function and gravitational potential, and apply only minimal physical assumptions.

\section{Method}

\label{sec:method}

Our first assumption is that stars orbit in a time-independent gravitational potential, $\Phi \left( \vec{x} \right)$. The density of an ensemble of stars in six-dimensional phase space (position $\vec{x}$ and velocity $\vec{v}$) is referred to as the \textit{distribution function}, $f \! \left( \vec{x} , \vec{v} \right)$. Liouville's theorem states that the total derivative of the distribution function of a collisionless system (in which the stars are not scattered by close-range interactions) is zero. For particles orbiting in a gravitational potential, this implies that
\begin{align}
  \frac{\mathrm{d}f}{\mathrm{d}t}
  =
  \frac{\partial f}{\partial t}
  \ + \!\!\!\!
  \sum_{\mathrm{dimension}\ i} \!
  \left(
    v_i \, \frac{\partial f}{\partial x_i}
    -\frac{\partial \Phi}{\partial x_i} \frac{\partial f}{\partial v_i}
  \right)
  = 0 \, .
  \label{eqn:f-total-derivative}
\end{align}
This is known as the ``collisionless Boltzmann equation.'' Our second assumption, that the distribution function is stationary, implies that the density in any given region of phase space is constant in time: $\frac{\partial f}{\partial t} = 0$. This assumption links gradients of the distribution function to gradients of the gravitational potential. \textit{Once we can describe the distribution function of a stationary system, in almost all cases, the gravitational potential can be uniquely determined by solving the collisionless Boltzmann equation} (Eq.~\ref{eqn:f-total-derivative}). Note that we do not assume that the gravitational potential is sourced by the observed stellar population alone. Accordingly, we do not impose $\nabla^2 \Phi = 4 \pi G \int \!f \left( \vec{x} , \vec{v} \right) \mathrm{d}^3\vec{v}$. Additional mass components, such as unobserved stars or dark matter, also contribute to the gravitational field.

\begin{figure}
  \centering
  \includegraphics[width=0.8\linewidth]{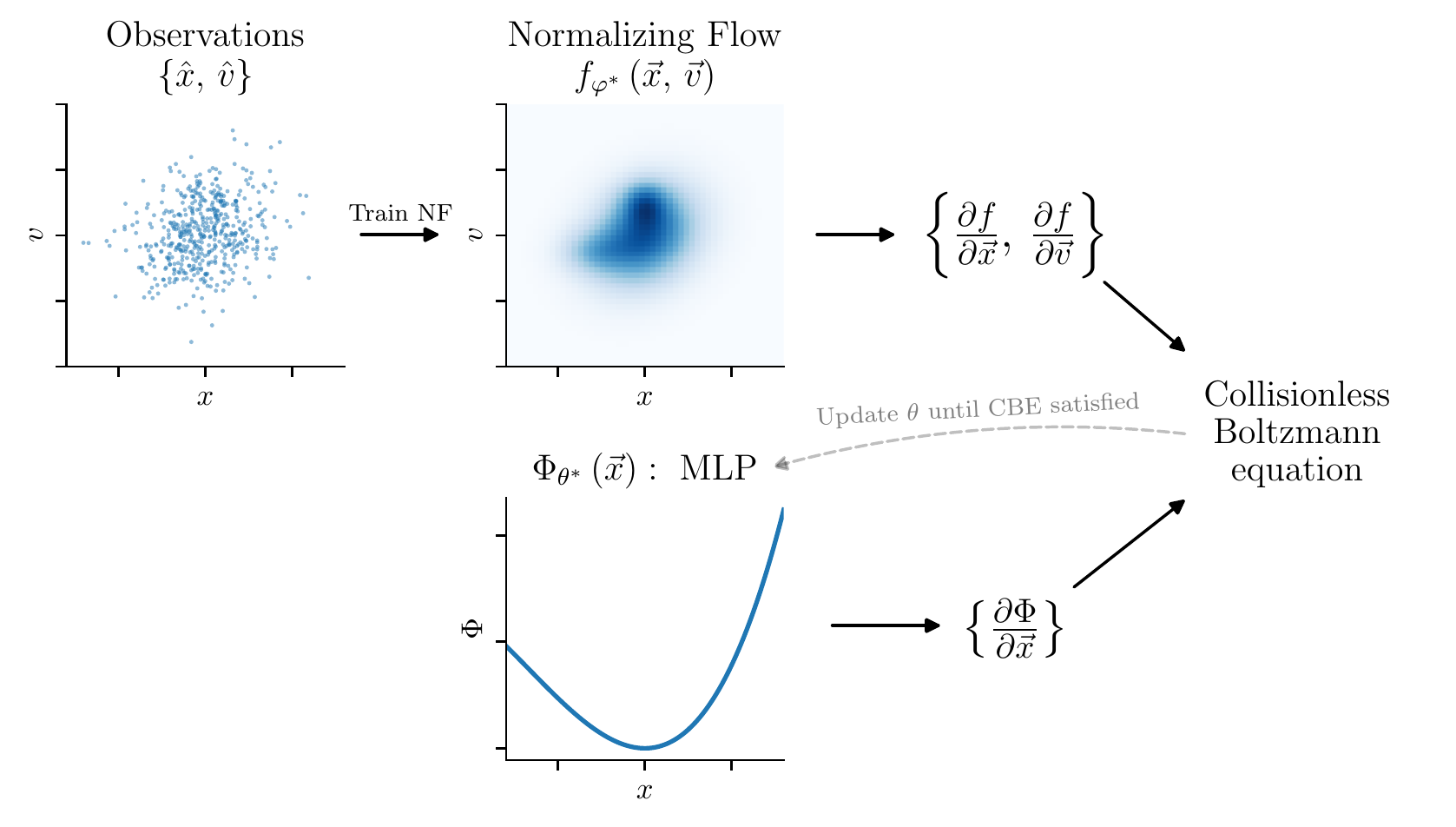}
  \caption{Overview of our method. Using observed phase-space information, we train a normalizing flow to represent the distribution function, $f \left( \vec{x} ,\, \vec{v} \right)$. We represent the gravitational potential, $\Phi \left( \vec{x} \right)$, by a feed-forward neural network. We update the neural network until the gradients of the potential and the distribution function satisfy the collisionless Boltzmann equation for a stationary system.}
  \label{fig:overview}
\end{figure}

In practice, when we observe stellar populations, we obtain a discrete sample of points in phase space, which we will refer to as $\left\{ \hat{x} , \hat{v} \right\}$. We do not directly observe the smooth distribution function, $f \left( \vec{x} , \vec{v} \right)$. A typical way of surmounting this difficulty is to fit a simple parametric model of the distribution function to the observed sample of phase-space locations of stars (e.g., \citep{BovyRix2013,McMillanBinney2013,Piffl2014,Trick2016}). Here, we instead represent the distribution function as a normalizing flow, trained on an ensemble of stars with measured phase-space information. We refer to the normalizing flow as $f_{\varphi} \! \left( \vec{x} , \vec{v} \, \right)$, where $\varphi$ refers to the trainable parameters in the flow. The best-fit flow parameters are those that maximize the Poisson likelihood of the phase-space coordinates of the stars:
\begin{align}
  \varphi^{\ast}
  &=
  \argmax_{\varphi} \bigg[
    \ln p \left( \left\{ \hat{x} , \hat{v} \right\} \mid \varphi \right)
  \bigg]
  =
  \argmax_{\varphi}
  \Bigg[
    \sum_{\mathrm{star}\ k}
      \ln f_{\varphi} \! \left( \hat{x}_k , \hat{v}_k \right)
  \Bigg]
  \, .
  \label{eqn:df-best-fit}
\end{align}
We thus obtain an approximation to the distribution function, $f_{\varphi^{\ast}}$. The great advantage of using a normalizing flow is that our representation is both highly flexible and auto-differentiable.

After learning the distribution function, we find the gravitational potential $\Phi \left( \vec{x} \right)$ that renders the distribution function stationary. The distribution is stationary if the sum in Eq.~\eqref{eqn:f-total-derivative} evaluates to zero everywhere in phase space. We also require that the matter density be non-negative everywhere in space. By Poisson's equation, which links the potential to the density, this implies that $\nabla^2 \Phi \geq 0$. We parameterize the gravitational potential as a feed-forward neural network, which takes a 3-vector, $\vec{x}$, and returns a scalar, $\Phi$. We denote the trainable parameters of this network by $\theta$, and the resulting approximation function as $\Phi_{\theta} \! \left( \vec{x} \right)$. We design a loss function that penalizes both non-stationarity and negative mass at phase-space points drawn from our approximation to the distribution function:
\begin{align}
  \theta^{\ast}
  &=
  \argmin_{\theta}
  \left<
    \asinh \left|
      \frac{\partial f_{\varphi^{\ast}}}{\partial t}
    \right|
    +
    \lambda \, \asinh \left(
      \max \left\{
        -\nabla^2 \Phi_{\theta}
        , \,
        0
      \right\}
    \right)
  \right>_{
    \vec{x} , \vec{v} \, \sim \, f_{\varphi^{\ast}}
  }
  \, ,
  \label{eqn:phi-best-fit}
\end{align}
where $\lambda$ is a hyperparameter that controls the relative weight given to the stationarity and positive-mass conditions ($\lambda = 2$ in this work), $\nicefrac{\partial f_{\varphi^{\ast}}}{\partial t}$ is calculated by plugging $\nicefrac{\partial \Phi_{\theta}}{\partial \vec{x}}$, $\nicefrac{\partial f_{\varphi^{\ast}}}{\partial \vec{x}}$ and $\nicefrac{\partial f_{\varphi^{\ast}}}{\partial \vec{v}}$ into Eq.~\eqref{eqn:f-total-derivative}. The $\asinh$ function is linear for small values, but lessens the influence of large outliers.

In the following demonstration, we use a normalizing flow architecture similar to \texttt{Glow} \citep{KingmaDhariwal2018}, but using rational-quadratic-spline coupling transforms \citep{Durkan2019} instead of affine coupling layers, and using an independent rescaling of each axis instead of \texttt{ActNorm} layers. One step of our flow architecture thus consists of an independent rescaling of each axis, an invertible $1 \! \times \! 1$ convolution and a rational-quadratic-spline coupling transform. Our flow architecture contains four steps. We find that we more accurately recover the distribution function with an \textit{ensemble} of 960 normalizing flows trained on identical data than with a single normalizing flow. The feed-forward neural network with which we represent the gravitational potential has three densely connected hidden layers, each with 128 neurons, and produces a scalar output. We implement these models in PyTorch,\footnote{\url{https://github.com/tingyuansen/deep-potential}. A Tensorflow 2.3 implementation can be found at \url{https://github.com/gregreen/deep-potential}.} and obtain $\varphi^{\ast}$ and $\theta^{\ast}$ through stochastic gradient descent, using the Rectified Adam optimizer \citep{Liu2019RAdam}.

\section{Demonstration}

\begin{figure}
  \centering
  \includegraphics[width=0.8\linewidth]{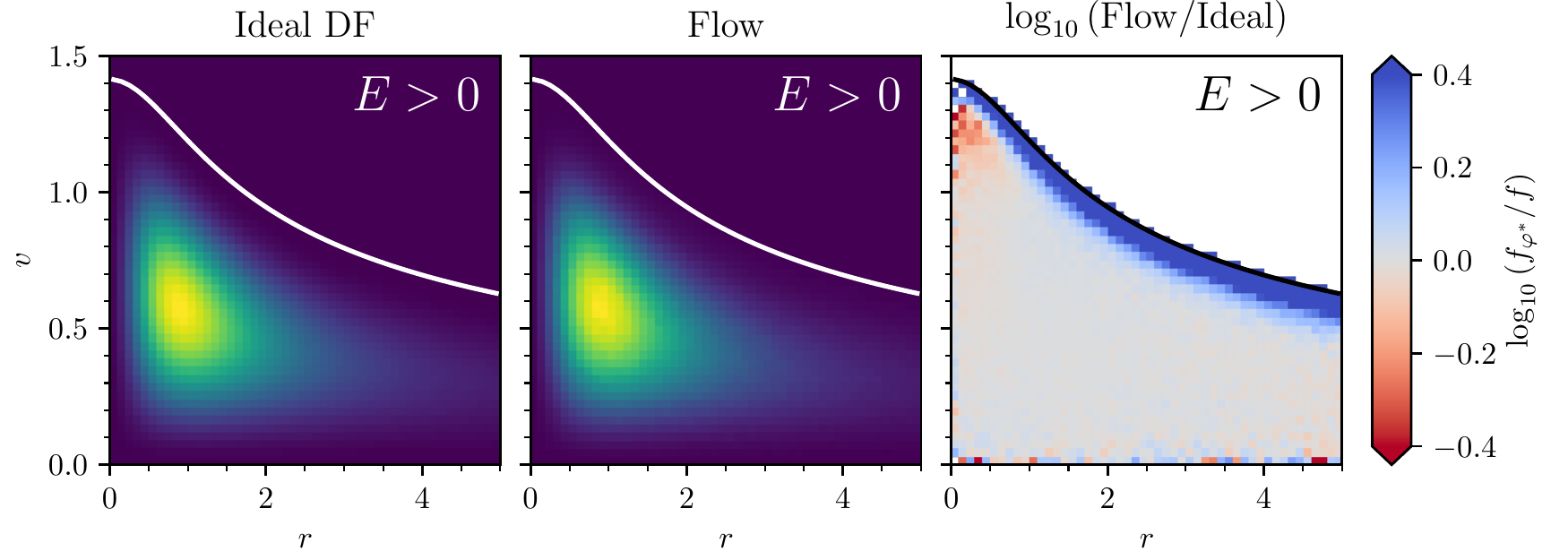}
  \caption{The ideal Plummer sphere distribution function (left panel), our trained ensemble of normalizing flows (middle panel), and a comparison of the two (right panel). We depict phase space in terms of radius and velocity, integrating over the four angular dimensions. Our ensemble of normalizing flows performs well in regions of non-negligible density.}
  \label{fig:flow}
\end{figure}

\begin{figure}
  \centering
  \includegraphics[width=0.8\linewidth]{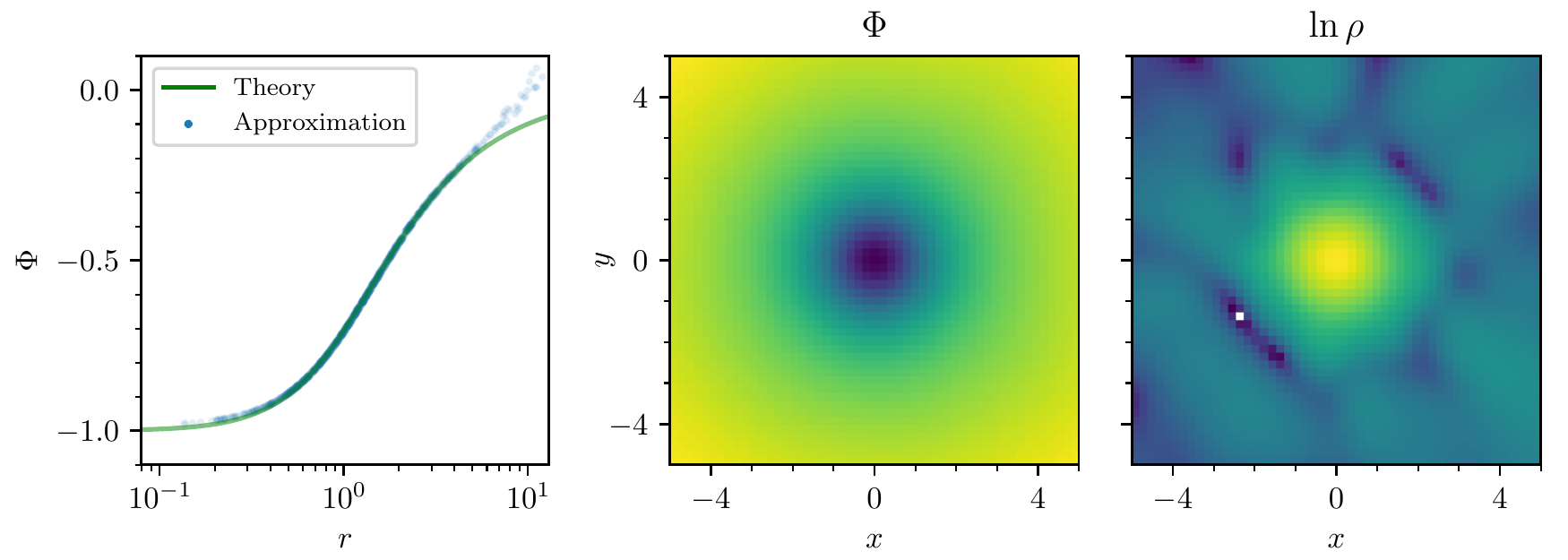}
  \caption{Left panel: comparison of the theoretical Plummer sphere potential with our result at random points drawn from phase space. The middle and right panels show our recovered potential and matter density in a 2D slice of space with $z = 0$.}
  \label{fig:potential}
\end{figure}

We demonstrate our method on a toy physical system, in which both the potential and distribution function can be expressed analytically. This constitutes a ground truth on which we can verify our method. We choose the Plummer sphere, a self-gravitating system with a spherically symmetric density and gravitational potential, given by
\begin{align}
    \rho \left( r \right) = \frac{3}{4\pi} \left( 1 + r^2 \right)^{-\nicefrac{5}{2}}
    \, ,
    \hspace{1cm}
    \Phi \left( r \right) = -\left( 1 + r^2 \right)^{-\nicefrac{1}{2}}
    \, .
\end{align}
The Plummer sphere admits a stationary distribution function with an isotropic velocity distribution, in which the distribution function is only a function of the energy $E$ of the tracer particle (for simplicity, we set mass $m = 1$ for all the particles):
\begin{align}
    f \! \left( \vec{r}, \vec{v} \right)
    &\propto
    \begin{cases}
      \left[ -E \left( \vec{r}, \vec{v} \right) \right]^{\nicefrac{7}{2}}, & E < 0 \\
      0, & E \geq 0
    \end{cases}
    \, , \hspace{1cm} \mathrm{where} \ 
    E = \frac{1}{2} v^2 + \Phi \left( r \right) \, .
\end{align}
We generate mock data by sampling $2^{17}$ (131,072) phase-space points drawn from the above distribution function. Using these points as input data, we first fit the distribution function using an ensemble of 960 normalizing flows (see Eq.~\ref{eqn:df-best-fit}). Our results are shown in Fig.~\ref{fig:flow}. Using our ensemble of flows, we then draw $2^{15}$ (32,768) phase-space points, and calculate the gradients $\nicefrac{\partial f_{\varphi^{\ast}}}{\partial \vec{x}}$, $\nicefrac{\partial f_{\varphi^{\ast}}}{\partial \vec{v}}$. We use these samples to fit the gravitational potential (see Eq.~\ref{eqn:phi-best-fit}). Our results are shown in Fig.~\ref{fig:potential}. We accurately recover the potential over a wide range of radii.

\section{Conclusions}

In this paper, we have shown that it is possible to accurately recover the gravitational potential of a stationary system using a snapshot of a sample of phase-mixed tracers. Auto-differentiable tensor frameworks are ideal tools for accomplishing this task, because of the need for smooth and differentiable -- yet highly flexible -- representations of the distribution function and gravitational potential. We have demonstrated that our method works on ideal mock data. Future work will examine how to take observational selection functions and errors into account. An additional avenue of research is how to apply our framework to physical systems that are not entirely phase-mixed, and therefore not completely stationary.

The \textit{Gaia} space telescope is currently surveying the parallaxes and proper motions of over a billion stars, and is additionally determining radial velocities of tens of millions of stars \citep{Prusti2016}. Ground-based spectroscopic surveys are set to deliver millions more high-precision radial velocities over the coming years \citep{Steinmetz2006,Zhao2012,Kollmeier2017}. We thus will soon have access to precise six-dimensional phase-space coordinates of tens of millions of stars throughout the Galaxy. The method that we present in this paper provides a means of extracting the gravitational potential -- and therefore the three-dimensional distribution of baryonic and dark matter in the Galaxy -- from these rich datasets, starting only from minimal physical assumptions (steady-state dynamics in a background gravitational field that corresponds to a positive matter density), and without resorting to restricted analytical models. This method therefore has the potential to reveal the full, unseen mass distribution of the Galaxy, using only a set of visible kinematic tracers -- the stars.

\section*{Broader Impact}

This work will advance our understanding of the dynamics of the Milky Way, and has the potential to uncover the distribution of both baryonic and dark matter throughout the Galaxy. The distribution of dark matter within a large spiral galaxy, such as the Milky Way, is of great interest to astronomers, particularly because it may help constrain models of the physical nature of dark matter. A great many people in the broader public are passionate about astronomy, and any advances in our understanding of the distribution of dark matter -- and anything we learn about the nature of dark matter from its spatial distribution -- will interest them. The authors do not foresee broader ethical concerns arising from this work.

\begin{ack}

Gregory Green acknowledges funding from the Alexander von Humboldt Foundation, through the Sofja Kovalevskaja Award.

Yuan-Sen Ting is grateful to be supported by the NASA Hubble Fellowship grant HST-HF2-51425.001 awarded by the Space Telescope Science Institute.

The authors thank Harshil Kamdar, Stephen Portillo, Hans-Walter Rix, Douglas Finkbeiner and Joshua Speagle for insightful discussions during the development of this method.

\end{ack}

\bibliographystyle{abbrvnat-maxnames-35}
\bibliography{neurips_2020}

\begin{thebibliography}{17}
\providecommand{\natexlab}[1]{#1}
\providecommand{\url}[1]{\texttt{#1}}
\expandafter\ifx\csname urlstyle\endcsname\relax
  \providecommand{\doi}[1]{doi: #1}\else
  \providecommand{\doi}{doi: \begingroup \urlstyle{rm}\Url}\fi

\bibitem[{Binney}(2013)]{Binney2013}
J.~{Binney}.
\newblock {Dynamics for galactic archaeology}.
\newblock \emph{New Astronomy Review}, 57\penalty0 (3-4):\penalty0 29--51,
  Sept. 2013.
\newblock \doi{10.1016/j.newar.2013.08.001}.

\bibitem[{Bland-Hawthorn} and {Gerhard}(2016)]{BlandHawthornGerhard2016}
J.~{Bland-Hawthorn} and O.~{Gerhard}.
\newblock {The Galaxy in Context: Structural, Kinematic, and Integrated
  Properties}.
\newblock \emph{Annual Review of Astronomy and Astrophysics}, 54:\penalty0
  529--596, Sept. 2016.
\newblock \doi{10.1146/annurev-astro-081915-023441}.

\bibitem[{Bovy} and {Rix}(2013)]{BovyRix2013}
J.~{Bovy} and H.-W. {Rix}.
\newblock {A Direct Dynamical Measurement of the Milky Way's Disk Surface
  Density Profile, Disk Scale Length, and Dark Matter Profile at 4 kpc < R < 9
  kpc}.
\newblock \emph{Astrophysical Journal}, 779\penalty0 (2):\penalty0 115, Dec.
  2013.
\newblock \doi{10.1088/0004-637X/779/2/115}.

\bibitem[{Durkan} et~al.(2019){Durkan}, {Bekasov}, {Murray}, and
  {Papamakarios}]{Durkan2019}
C.~{Durkan}, A.~{Bekasov}, I.~{Murray}, and G.~{Papamakarios}.
\newblock {Neural Spline Flows}.
\newblock \emph{arXiv e-prints}, art. arXiv:1906.04032, June 2019.

\bibitem[{Gaia Collaboration} et~al.(2016){Gaia Collaboration}, {Prusti}, {de
  Bruijne}, {Brown}, {Vallenari}, {Babusiaux}, {Bailer-Jones}, {Bastian},
  {Biermann}, {Evans}, {Eyer}, {Jansen}, {Jordi}, {Klioner}, {Lammers},
  {Lindegren}, {Luri}, {Mignard}, {Milligan}, {Panem}, {Poinsignon},
  {Pourbaix}, {Randich}, {Sarri}, {Sartoretti}, {Siddiqui}, {Soubiran},
  {Valette}, {van Leeuwen}, {Walton}, {Aerts}, {Arenou}, {Cropper}, {Drimmel},
  {H{\o}g}, {Katz}, {Lattanzi}, {O'Mullane}, {Grebel}, {Holland}, {Huc},
  {Passot}, {Bramante}, {Cacciari}, {Casta{\~n}eda}, {Chaoul}, {Cheek}, {De
  Angeli}, {Fabricius}, {Guerra}, {Hern{\'a}ndez}, {Jean-Antoine-Piccolo},
  {Masana}, {Messineo}, {Mowlavi}, {Nienartowicz}, {Ord{\'o}{\~n}ez-Blanco},
  {Panuzzo}, {Portell}, {Richards}, {Riello}, {Seabroke}, {Tanga},
  {Th{\'e}venin}, {Torra}, {Els}, {Gracia-Abril}, {Comoretto},
  {Garcia-Reinaldos}, {Lock}, {Mercier}, {Altmann}, {Andrae}, {Astraatmadja},
  {Bellas-Velidis}, {Benson}, {Berthier}, {Blomme}, {Busso}, {Carry},
  {Cellino}, {Clementini}, {Cowell}, {Creevey}, {Cuypers}, {Davidson}, {De
  Ridder}, {de Torres}, {Delchambre}, {Dell'Oro}, {Ducourant}, {Fr{\'e}mat},
  {Garc{\'\i}a-Torres}, {Gosset}, {Halbwachs}, {Hambly}, {Harrison}, {Hauser},
  {Hestroffer}, {Hodgkin}, {Huckle}, {Hutton}, {Jasniewicz}, {Jordan},
  {Kontizas}, {Korn}, {Lanzafame}, {Manteiga}, {Moitinho}, {Muinonen},
  {Osinde}, {Pancino}, {Pauwels}, {Petit}, {Recio-Blanco}, {Robin}, {Sarro},
  {Siopis}, {Smith}, {Smith}, {Sozzetti}, {Thuillot}, {van Reeven}, {Viala},
  {Abbas}, {Abreu Aramburu}, {Accart}, {Aguado}, {Allan}, {Allasia},
  {Altavilla}, {{\'A}lvarez}, {Alves}, {Anderson}, {Andrei}, {Anglada Varela},
  {Antiche}, {Antoja}, {Ant{\'o}n}, {Arcay}, {Atzei}, {Ayache}, {Bach},
  {Baker}, {Balaguer-N{\'u}{\~n}ez}, {Barache}, {Barata}, {Barbier}, {Barblan},
  {Baroni}, {Barrado y Navascu{\'e}s}, {Barros}, {Barstow}, {Becciani},
  {Bellazzini}, {Bellei}, {Bello Garc{\'\i}a}, {Belokurov}, {Bendjoya},
  {Berihuete}, {Bianchi}, {Bienaym{\'e}}, {Billebaud}, {Blagorodnova},
  {Blanco-Cuaresma}, {Boch}, {Bombrun}, {Borrachero}, {Bouquillon}, {Bourda},
  {Bouy}, {Bragaglia}, {Breddels}, {Brouillet}, {Br{\"u}semeister},
  {Bucciarelli}, {Budnik}, {Burgess}, {Burgon}, {Burlacu}, {Busonero}, {Buzzi},
  {Caffau}, {Cambras}, {Campbell}, {Cancelliere}, {Cantat-Gaudin}, {Carlucci},
  {Carrasco}, {Castellani}, {Charlot}, {Charnas}, {Charvet}, {Chassat},
  {Chiavassa}, {Clotet}, {Cocozza}, {Collins}, {Collins}, {Costigan}, {Crifo},
  {Cross}, {Crosta}, {Crowley}, {Dafonte}, {Damerdji}, {Dapergolas}, {David},
  {David}, {De Cat}, {de Felice}, {de Laverny}, {De Luise}, {De March}, {de
  Martino}, {de Souza}, {Debosscher}, {del Pozo}, {Delbo}, {Delgado},
  {Delgado}, {di Marco}, {Di Matteo}, {Diakite}, {Distefano}, {Dolding}, {Dos
  Anjos}, {Drazinos}, {Dur{\'a}n}, {Dzigan}, {Ecale}, {Edvardsson}, {Enke},
  {Erdmann}, {Escolar}, {Espina}, {Evans}, {Eynard Bontemps}, {Fabre},
  {Fabrizio}, {Faigler}, {Falc{\~a}o}, {Farr{\`a}s Casas}, {Faye}, {Federici},
  {Fedorets}, {Fern{\'a}ndez-Hern{\'a}ndez}, {Fernique}, {Fienga}, {Figueras},
  {Filippi}, {Findeisen}, {Fonti}, {Fouesneau}, {Fraile}, {Fraser}, {Fuchs},
  {Furnell}, {Gai}, {Galleti}, {Galluccio}, {Garabato}, {Garc{\'\i}a-Sedano},
  {Gar{\'e}}, {Garofalo}, {Garralda}, {Gavras}, {Gerssen}, {Geyer}, {Gilmore},
  {Girona}, {Giuffrida}, {Gomes}, {Gonz{\'a}lez-Marcos},
  {Gonz{\'a}lez-N{\'u}{\~n}ez}, {Gonz{\'a}lez-Vidal}, {Granvik}, {Guerrier},
  {Guillout}, {Guiraud}, {G{\'u}rpide}, {Guti{\'e}rrez-S{\'a}nchez}, {Guy},
  {Haigron}, {Hatzidimitriou}, {Haywood}, {Heiter}, {Helmi}, {Hobbs},
  {Hofmann}, {Holl}, {Holland }, {Hunt}, {Hypki}, {Icardi}, {Irwin}, {Jevardat
  de Fombelle}, {Jofr{\'e}}, {Jonker}, {Jorissen}, {Julbe}, {Karampelas},
  {Kochoska}, {Kohley}, {Kolenberg}, {Kontizas}, {Koposov}, {Kordopatis},
  {Koubsky}, {Kowalczyk}, {Krone-Martins}, {Kudryashova}, {Kull}, {Bachchan},
  {Lacoste-Seris}, {Lanza}, {Lavigne}, {Le Poncin-Lafitte}, {Lebreton},
  {Lebzelter}, {Leccia}, {Leclerc}, {Lecoeur-Taibi}, {Lemaitre}, {Lenhardt},
  {Leroux}, {Liao}, {Licata}, {Lindstr{\o}m}, {Lister}, {Livanou}, {Lobel},
  {L{\"o}ffler}, {L{\'o}pez}, {Lopez-Lozano}, {Lorenz}, {Loureiro},
  {MacDonald}, {Magalh{\~a}es Fernandes}, {Managau}, {Mann}, {Mantelet},
  {Marchal}, {Marchant}, {Marconi}, {Marie}, {Marinoni}, {Marrese},
  {Marschalk{\'o}}, {Marshall}, {Mart{\'\i}n-Fleitas}, {Martino}, {Mary},
  {Matijevi{\v{c}}}, {Mazeh}, {McMillan}, {Messina}, {Mestre}, {Michalik},
  {Millar}, {Miranda}, {Molina}, {Molinaro}, {Molinaro}, {Moln{\'a}r},
  {Moniez}, {Montegriffo}, {Monteiro}, {Mor}, {Mora}, {Morbidelli}, {Morel},
  {Morgenthaler}, {Morley}, {Morris}, {Mulone}, {Muraveva}, {Musella},
  {Narbonne}, {Nelemans}, {Nicastro}, {Noval}, {Ord{\'e}novic},
  {Ordieres-Mer{\'e}}, {Osborne}, {Pagani}, {Pagano}, {Pailler}, {Palacin},
  {Palaversa}, {Parsons}, {Paulsen}, {Pecoraro}, {Pedrosa}, {Pentik{\"a}inen},
  {Pereira}, {Pichon}, {Piersimoni}, {Pineau}, {Plachy}, {Plum}, {Poujoulet},
  {Pr{\v{s}}a}, {Pulone}, {Ragaini}, {Rago}, {Rambaux}, {Ramos-Lerate},
  {Ranalli}, {Rauw}, {Read}, {Regibo}, {Renk}, {Reyl{\'e}}, {Ribeiro},
  {Rimoldini}, {Ripepi}, {Riva}, {Rixon}, {Roelens}, {Romero-G{\'o}mez},
  {Rowell}, {Royer}, {Rudolph}, {Ruiz-Dern}, {Sadowski}, {Sagrist{\`a}
  Sell{\'e}s}, {Sahlmann}, {Salgado}, {Salguero}, {Sarasso}, {Savietto},
  {Schnorhk}, {Schultheis}, {Sciacca}, {Segol}, {Segovia}, {Segransan},
  {Serpell}, {Shih}, {Smareglia}, {Smart}, {Smith}, {Solano}, {Solitro},
  {Sordo}, {Soria Nieto}, {Souchay}, {Spagna}, {Spoto}, {Stampa}, {Steele},
  {Steidelm{\"u}ller}, {Stephenson}, {Stoev}, {Suess}, {S{\"u}veges}, {Surdej},
  {Szabados}, {Szegedi-Elek}, {Tapiador}, {Taris}, {Tauran}, {Taylor},
  {Teixeira}, {Terrett}, {Tingley}, {Trager}, {Turon}, {Ulla}, {Utrilla},
  {Valentini}, {van Elteren}, {Van Hemelryck}, {van Leeuwen}, {Varadi},
  {Vecchiato}, {Veljanoski}, {Via}, {Vicente}, {Vogt}, {Voss}, {Votruba},
  {Voutsinas}, {Walmsley}, {Weiler}, {Weingrill}, {Werner}, {Wevers},
  {Whitehead}, {Wyrzykowski}, {Yoldas}, {{\v{Z}}erjal}, {Zucker}, {Zurbach},
  {Zwitter}, {Alecu}, {Allen}, {Allende Prieto}, {Amorim},
  {Anglada-Escud{\'e}}, {Arsenijevic}, {Azaz}, {Balm}, {Beck}, {Bernstein},
  {Bigot}, {Bijaoui}, {Blasco}, {Bonfigli}, {Bono}, {Boudreault}, {Bressan},
  {Brown}, {Brunet}, {Bunclark}, {Buonanno}, {Butkevich}, {Carret}, {Carrion},
  {Chemin}, {Ch{\'e}reau}, {Corcione}, {Darmigny}, {de Boer}, {de Teodoro}, {de
  Zeeuw}, {Delle Luche}, {Domingues}, {Dubath}, {Fodor}, {Fr{\'e}zouls},
  {Fries}, {Fustes}, {Fyfe}, {Gallardo}, {Gallegos}, {Gardiol}, {Gebran},
  {Gomboc}, {G{\'o}mez}, {Grux}, {Gueguen}, {Heyrovsky}, {Hoar}, {Iannicola},
  {Isasi Parache}, {Janotto}, {Joliet}, {Jonckheere}, {Keil}, {Kim},
  {Klagyivik}, {Klar}, {Knude}, {Kochukhov}, {Kolka}, {Kos}, {Kutka}, {Lainey},
  {LeBouquin}, {Liu}, {Loreggia}, {Makarov}, {Marseille}, {Martayan},
  {Martinez-Rubi}, {Massart}, {Meynadier}, {Mignot}, {Munari}, {Nguyen},
  {Nordlander}, {Ocvirk}, {O'Flaherty}, {Olias Sanz}, {Ortiz}, {Osorio},
  {Oszkiewicz}, {Ouzounis}, {Palmer}, {Park}, {Pasquato}, {Peltzer}, {Peralta},
  {P{\'e}turaud}, {Pieniluoma}, {Pigozzi}, {Poels}, {Prat}, {Prod'homme},
  {Raison}, {Rebordao}, {Risquez}, {Rocca-Volmerange}, {Rosen}, {Ruiz-Fuertes},
  {Russo}, {Sembay}, {Serraller Vizcaino}, {Short}, {Siebert}, {Silva},
  {Sinachopoulos}, {Slezak}, {Soffel}, {Sosnowska}, {Strai{\v{z}}ys}, {ter
  Linden}, {Terrell}, {Theil}, {Tiede}, {Troisi}, {Tsalmantza}, {Tur},
  {Vaccari}, {Vachier}, {Valles}, {Van Hamme}, {Veltz}, {Virtanen}, {Wallut},
  {Wichmann}, {Wilkinson}, {Ziaeepour}, and {Zschocke}]{Prusti2016}
{Gaia Collaboration}, T.~{Prusti}, J.~H.~J. {de Bruijne}, et~al.
\newblock {The Gaia mission}.
\newblock \emph{Astronomy and Astrophysics}, 595:\penalty0 A1, Nov. 2016.
\newblock \doi{10.1051/0004-6361/201629272}.

\bibitem[{Greydanus} et~al.(2019){Greydanus}, {Dzamba}, and
  {Yosinski}]{Greydanus2019}
S.~{Greydanus}, M.~{Dzamba}, and J.~{Yosinski}.
\newblock {Hamiltonian Neural Networks}.
\newblock \emph{arXiv e-prints}, art. arXiv:1906.01563, June 2019.

\bibitem[{Kingma} and {Dhariwal}(2018)]{KingmaDhariwal2018}
D.~P. {Kingma} and P.~{Dhariwal}.
\newblock {Glow: Generative Flow with Invertible 1x1 Convolutions}.
\newblock \emph{arXiv e-prints}, art. arXiv:1807.03039, July 2018.

\bibitem[{Kollmeier} et~al.(2017){Kollmeier}, {Zasowski}, {Rix}, {Johns},
  {Anderson}, {Drory}, {Johnson}, {Pogge}, {Bird}, {Blanc}, {Brownstein},
  {Crane}, {De Lee}, {Klaene}, {Kreckel}, {MacDonald}, {Merloni}, {Ness},
  {O'Brien}, {Sanchez-Gallego}, {Sayres}, {Shen}, {Thakar}, {Tkachenko},
  {Aerts}, {Blanton}, {Eisenstein}, {Holtzman}, {Maoz}, {Nandra}, {Rockosi},
  {Weinberg}, {Bovy}, {Casey}, {Chaname}, {Clerc}, {Conroy}, {Eracleous},
  {G{\"a}nsicke}, {Hekker}, {Horne}, {Kauffmann}, {McQuinn}, {Pellegrini},
  {Schinnerer}, {Schlafly}, {Schwope}, {Seibert}, {Teske}, and {van
  Saders}]{Kollmeier2017}
J.~A. {Kollmeier}, G.~{Zasowski}, H.-W. {Rix}, et~al.
\newblock {SDSS-V: Pioneering Panoptic Spectroscopy}.
\newblock \emph{arXiv e-prints}, art. arXiv:1711.03234, Nov. 2017.

\bibitem[{Liu} et~al.(2019){Liu}, {Jiang}, {He}, {Chen}, {Liu}, {Gao}, and
  {Han}]{Liu2019RAdam}
L.~{Liu}, H.~{Jiang}, P.~{He}, et~al.
\newblock {On the Variance of the Adaptive Learning Rate and Beyond}.
\newblock \emph{arXiv e-prints}, art. arXiv:1908.03265, Aug. 2019.

\bibitem[{McMillan} and {Binney}(2013)]{McMillanBinney2013}
P.~J. {McMillan} and J.~J. {Binney}.
\newblock {Analysing surveys of our Galaxy - II. Determining the potential}.
\newblock \emph{Monthly Notices of the Royal Astronomical Society},
  433\penalty0 (2):\penalty0 1411--1424, Aug. 2013.
\newblock \doi{10.1093/mnras/stt814}.

\bibitem[{Piffl} et~al.(2014){Piffl}, {Binney}, {McMillan}, {Steinmetz},
  {Helmi}, {Wyse}, {Bienaym{\'e}}, {Bland -Hawthorn}, {Freeman}, {Gibson},
  {Gilmore}, {Grebel}, {Kordopatis}, {Navarro}, {Parker}, {Reid}, {Seabroke},
  {Siebert}, {Watson}, and {Zwitter}]{Piffl2014}
T.~{Piffl}, J.~{Binney}, P.~J. {McMillan}, et~al.
\newblock {Constraining the Galaxy's dark halo with RAVE stars}.
\newblock \emph{Monthly Notices of the Royal Astronomical Society},
  445\penalty0 (3):\penalty0 3133--3151, Dec. 2014.
\newblock \doi{10.1093/mnras/stu1948}.

\bibitem[{Quercellini} et~al.(2008){Quercellini}, {Amendola}, and
  {Balbi}]{Quercellini2008}
C.~{Quercellini}, L.~{Amendola}, and A.~{Balbi}.
\newblock {Mapping the galactic gravitational potential with peculiar
  acceleration}.
\newblock \emph{Monthly Notices of the Royal Astronomical Society},
  391\penalty0 (3):\penalty0 1308--1314, Dec. 2008.
\newblock \doi{10.1111/j.1365-2966.2008.13968.x}.

\bibitem[{Ravi} et~al.(2019){Ravi}, {Langellier}, {Phillips}, {Buschmann},
  {Safdi}, and {Walsworth}]{Ravi2019}
A.~{Ravi}, N.~{Langellier}, D.~F. {Phillips}, et~al.
\newblock {Probing Dark Matter Using Precision Measurements of Stellar
  Accelerations}.
\newblock \emph{Physical Review Letters}, 123\penalty0 (9):\penalty0 091101,
  Aug. 2019.
\newblock \doi{10.1103/PhysRevLett.123.091101}.

\bibitem[{Silverwood} and {Easther}(2019)]{Silverwood2019}
H.~{Silverwood} and R.~{Easther}.
\newblock {Stellar accelerations and the galactic gravitational field}.
\newblock \emph{Publications of the Astronomical Society of Australia},
  36:\penalty0 e038, Oct. 2019.
\newblock \doi{10.1017/pasa.2019.25}.

\bibitem[{Steinmetz} et~al.(2006){Steinmetz}, {Zwitter}, {Siebert}, {Watson},
  {Freeman}, {Munari}, {Campbell}, {Williams}, {Seabroke}, {Wyse}, {Parker},
  {Bienaym{\'e}}, {Roeser}, {Gibson}, {Gilmore}, {Grebel}, {Helmi}, {Navarro},
  {Burton}, {Cass}, {Dawe}, {Fiegert}, {Hartley}, {Russell}, {Saunders},
  {Enke}, {Bailin}, {Binney}, {Bland -Hawthorn}, {Boeche}, {Dehnen},
  {Eisenstein}, {Evans}, {Fiorucci}, {Fulbright}, {Gerhard}, {Jauregi}, {Kelz},
  {Mijovi{\'c}}, {Minchev}, {Parmentier}, {Pe{\~n}arrubia}, {Quillen}, {Read},
  {Ruchti}, {Scholz}, {Siviero}, {Smith}, {Sordo}, {Veltz}, {Vidrih}, {von
  Berlepsch}, {Boyle}, and {Schilbach}]{Steinmetz2006}
M.~{Steinmetz}, T.~{Zwitter}, A.~{Siebert}, et~al.
\newblock {The Radial Velocity Experiment (RAVE): First Data Release}.
\newblock \emph{Astronomical Journal}, 132\penalty0 (4):\penalty0 1645--1668,
  Oct. 2006.
\newblock \doi{10.1086/506564}.

\bibitem[{Trick} et~al.(2016){Trick}, {Bovy}, and {Rix}]{Trick2016}
W.~H. {Trick}, J.~{Bovy}, and H.-W. {Rix}.
\newblock {Action-Based Dynamical Modeling for the Milky Way Disk}.
\newblock \emph{Astrophysical Journal}, 830\penalty0 (2):\penalty0 97, Oct.
  2016.
\newblock \doi{10.3847/0004-637X/830/2/97}.

\bibitem[{Zhao} et~al.(2012){Zhao}, {Zhao}, {Chu}, {Jing}, and
  {Deng}]{Zhao2012}
G.~{Zhao}, Y.-H. {Zhao}, Y.-Q. {Chu}, et~al.
\newblock {LAMOST spectral survey {\textemdash} An overview}.
\newblock \emph{Research in Astronomy and Astrophysics}, 12\penalty0
  (7):\penalty0 723--734, July 2012.
\newblock \doi{10.1088/1674-4527/12/7/002}.

\end{thebibliography}

\end{document}